\documentclass[smallextended]{svjour2}
\smartqed
\usepackage{dcolumn}
\usepackage{bm}
\usepackage{graphicx}
\usepackage{hyperref}
\usepackage{mathptmx}
\usepackage{subfigure}
\usepackage[utf8]{inputenc}
\usepackage{bbold}
\usepackage{color}
\usepackage{amssymb}

\usepackage{amsmath}

\newcommand{\OO}{\mathcal{O}_L}
\newcommand{\CC}{\mathcal{C}_L}
\newcommand{\CCR}{\mathcal{C}_{r,L}}
\newcommand{\CCC}{\mathcal{C}_{\mathrm{co},L}}

\newcommand{\OOp}[1]{\mathcal{O}_L(#1)}
\newcommand{\CCp}[1]{\mathcal{C}_L(#1)}

\newcommand{\Pop}{P_{\mathrm{o}, L}}
\newcommand{\Pc}{P_{\mathrm{c}, L}}
\newcommand{\Pv}[2]{P_{\mathrm{#1#2}, L}}

\newcommand{\rr}{r}

\newcommand{\lco}{\ell_{\mathrm{co}, L}}
\newcommand{\lc}{\ell_{\mathrm{c}, L}}

\newcommand{\Nc}{N_{\mathrm{co},L}}

\newcommand{\pp}{p}

\makeatletter

\begin{document}

\title{Zero density of open paths in the Lorentz mirror model for arbitrary mirror probability}
\author{Atahualpa S.~Kraemer \and David P.~Sanders}

\institute{Departamento de F\'isica, Facultad de Ciencias, Universidad Nacional
Aut\'onoma de M\'exico,
Ciudad Universitaria, M\'exico D.F.\ 04510, Mexico
}

\date{\today}

\maketitle
%

\email{ata.kraemer@gmail.com}
\email{dpsanders@ciencias.unam.mx}


\begin{abstract}
We show, incorporating results obtained from numerical simulations, that in the Lorentz mirror model,
the density of open paths in any finite box tends to $0$ as the box size tends to
infinity, for any mirror probability.
\end{abstract}

\section{Introduction}

Lorentz gases are models that consist of particles moving in space 
and colliding elastically with spherical obstacles
placed on the vertices of a grid, which can be periodic (see, e.g., \cite{1,6,7}) or aperiodic
(see, e.g., \cite{5,ata-david}). A
similar model is the Ehrenfest wind--tree model, in which the obstacles are
diamonds \cite{164}. 

A version of the Ehrenfest model is to consider particles moving along the edges of a
lattice,
placing obstacles at the vertices that deflect particles according to a
deterministic rule; such models are known as lattice Lorentz  gases. Interest has focused
on their diffusive properties 
\cite{XP-Kong1,XP-Kong2,M-H-Ernst1,M-H-Ernst2,Cohen1,XP-Kong3,Wang2,Beijeren,Wang,Acedo,Cohen2}, the distribution of orbit sizes
\cite{ziff}, topological dynamics \cite{Bunimovich2}  and recurrence properties
\cite{10,Bunimovich,Quas,ziff,GADY-KOZMA}.

One of the most-studied versions of the lattice Lorentz gas is the Lorentz mirror model \cite{GADY-KOZMA},
introduced in the physics literature by Ruijgrok and Cohen~\cite{Ruijgrok}, where the scatterers are two-sided mirrors
placed on the sites of a square lattice, $\mathbb{Z}^2$. These mirrors are fixed, with an orientation of $45$ degrees
(right) or $-45$ degrees (left), and reflect particles (or light beams) that
 move along the edges of the lattice. The orientation of each mirror is
chosen uniformly at random, with probabilities $\pp_r$ and $\pp_l$, respectively, with $\pp := \pp_{l} + \pp_{r} \le 1$;  the mirror model is thus one of the simplest 
realisations of deterministic dynamics in a quenched random environment. 
(There is also a flipping version of the mirror model, in which the mirrors change orientation when hit \cite{Ruijgrok}.)
When a particle arrives at a site occupied by a mirror, it
is reflected, and leaves along one of the other three edges incident on that site, following the rule given by the mirror (or absence of a mirror) there.
Thus, either exactly two distinct paths touch a given mirror, or a single
path touches the mirror from all directions.

For this model, it 
was proved that  if the total density of mirrors
$\pp = \pp_r+ \pp_l$ is $1$, and both individual orientation probabilities are greater than $0$, then with probability $1$ (with respect to the random environment) all trajectories are periodic, forming a closed path, \cite{10,Bunimovich}. 

For concentrations $0 < \pp < 1$, it was conjectured \cite{181}, based, in part, on numerical simulations  \cite{10}, that the same holds for any mirror concentration $\pp$. However, we are not aware of any previous argument, even of a heuristic nature, to confirm or refute this conjecture. Of the few rigorous results known for $\pp < 1$, Quas proved that if there exist open orbits, then one of the following holds: There is only one open orbit, or there are infinitely many open
orbits \cite{Quas}. Recently, Kozma and Sidoravicius gave a rigorous lower bound on the probability that a trajectory from the origin leaves a square box of given side length $L$.

In this letter we show, incorporating numerical results in our argument, that a weaker result holds: for any mirror density $\pp > 0$, the \emph{density} of open paths (i.e., the fraction of edges which are contained in open paths) converges to $0$ in the limit $L \to \infty$.


\section{Definitions of key quantities}

We will work exclusively with \emph{finite} portions of the infinite lattice, consisting of 
square boxes with side length $L$:
 mirrors are placed on the $L^2$ vertices, and there are $2L^2$ edges (excluding those that leave the box).
The trajectory of a particle forms a \emph{path}, i.e., the directed sequence of edges
followed by the particle under the dynamics.
In this finite version of the model, some trajectories form \emph{closed paths} (periodic orbits),  as in the infinite version; others escape from the finite box, and we call these \emph{open paths}.  Note that open paths necessarily are open on both ends, crossing from one edge of the box to another (possibly the same edge). There are thus exactly 
$2L$ open paths.

The main quantities of interest in our analysis are the total lengths of all closed and open paths in the system of size $L$, i.e., the total number of \emph{edges} contained in all closed paths and all open paths, respectively; we denote their means (over all realisations of the disorder in a box of size $L$), as functions of the mirror density $p$, 
 by $\CCp{\pp}$ and $\OOp{\pp}$. For brevity, we denote $\OO := \OOp{1}$, and we similarly omit the argument for
other quantities which depend on the mirror density $p$ when $p=1$.
Since all edges belong either to an open path, or to a closed path, we have $\CCp {\pp}+
\OOp{\pp} = 2L^{2}$ for all $\pp$.

The theorem of Grimmett and Bunimovich and Troubetzkoy on the infinite model states that in the limit $L \to \infty$, with probability $1$ (with respect to the probability distribution [product measure] of the disorder), all trajectories close 
\cite{181,Bunimovich}.
This implies the weaker statement that  $\CC / 2L^{2} \rightarrow 1$ as $L \to \infty$, and hence
$\OO / 2L^{2} \rightarrow 0$. Our claim is that these also hold when $0 < p < 1$, i.e. that $\OOp{p} / 2L^{2} \rightarrow 0$, for any $0 < p \le 1$, when $L \to \infty$. This implies that no phase transition occurs -- this would correspond to the existence of a critical probability $p_{c}$ such that a positive fraction of paths are open below $p_{c}$.

\begin{figure}[t]
\begin{center}
\includegraphics[scale=0.8]{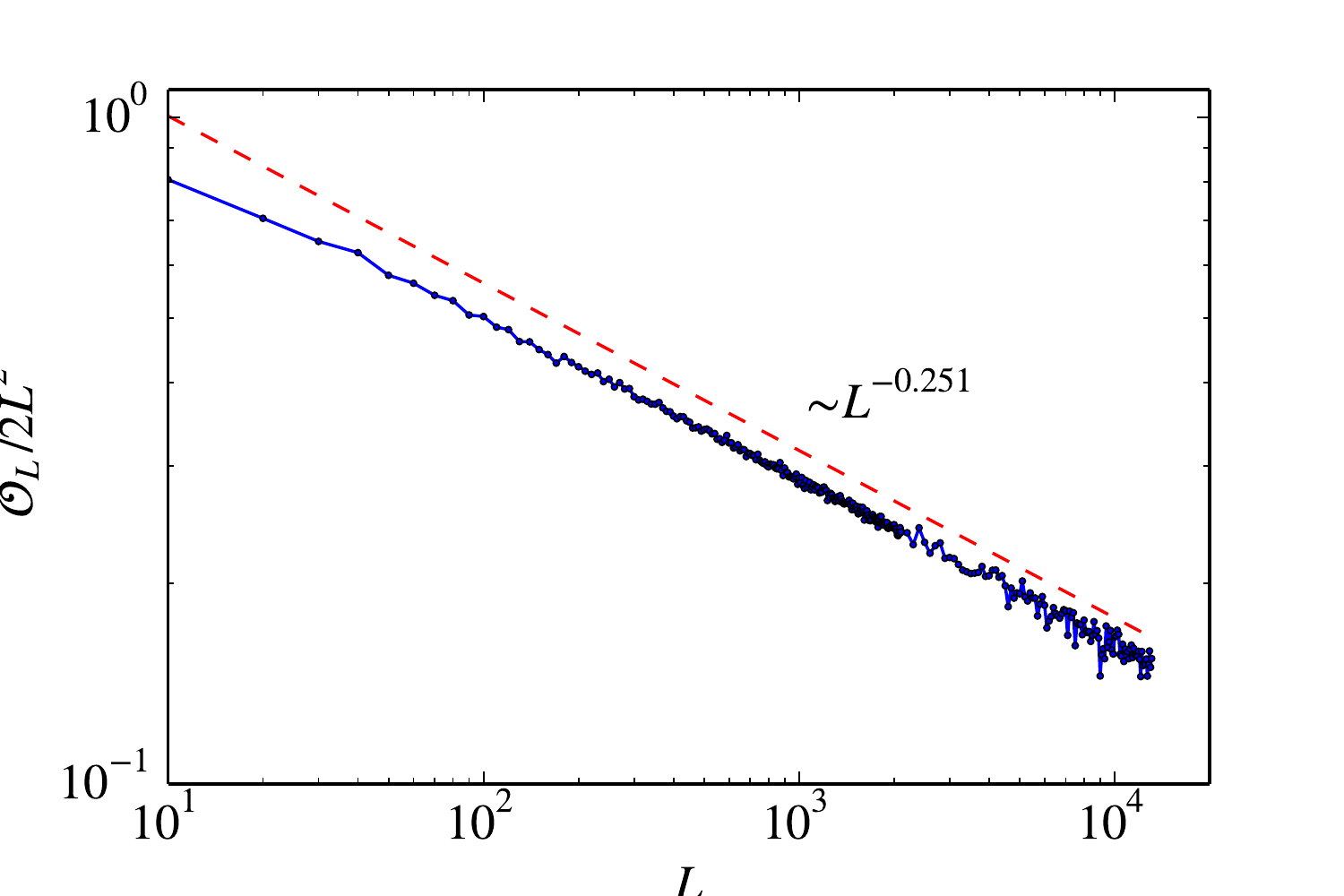}
\caption{Density $\OO / 2L^{2}$ of open paths as a function of box size $L$. The entire box of side length $L$ was simulated and each relevant quantity determined; each data point is averaged over at least 10 runs. The lines joining the points are a guide for the eye, and the dashed line shows a power-law fit to the second half of the data points.}
\label{fig:open_length}
\end{center}
\end{figure}

Our analysis uses the asymptotic behaviour of the density of open paths as $L \to \infty$ in a fundamental way.
Figure~\ref{fig:open_length} shows results of numerical simulations of this quantity. (Details of the efficient algorithms employed will be reported in a separate publication.)
 We find that the asymptotic behaviour is
  \begin{equation}
  \OO/ 2L^{2} \sim C \, L^{-\alpha} \qquad \mathrm{ with } \quad \alpha \simeq 0.251,
  \end{equation} 
  where $C$ is a constant.
We find that $\OO
\sim L^{2-\alpha} \sim L^{1.75}$; this exponent seems to be equal to the fractal
dimension $d_{f}  = \frac{7}{4}$ of the
trajectories \cite{ziff}. To explain this, consider those open paths that traverse the box 
from one side of the system to the other. These belong to a percolating cluster,
and are fractal with exponent $7/4$ \cite{Ziff2}. Many of the open paths
will be short, of length of order $L$, being localised close to the boundaries of the square box,
and thus will be dominated by the long paths that cross the system. This implies that
 $\OO$ will scale as $L^{7/4}$; dividing by $L^2$ gives the exponent $\alpha=-1/4$.

We denote by $\Pc:= \CC / 2L^{2}$  the probability that a given \emph{edge} belongs to a closed path, and
$\Pop := 1 - \Pc$  the probability that an edge belongs to an open path, both for
mirror density $\pp = 1$.
We also  denote by $\Pv{o}{o}$ the probability that a randomly-chosen
\emph{vertex}
separates two open paths (or a single open path with itself), and similarly for the other
combinations of open and closed. 

Counting the number of open edges (those which form part of an open path) incident on each type of  vertex gives the equality
\begin{equation}
2 \Pop = \textstyle \Pv{o}{c} + 2 \Pv{o}{o}.
\end{equation}

We will study the asymptotic behaviour of the above quantities as $L \to \infty$,
for which we will use the  notation $A_L \lesssim B_L$ to mean that
$\lim_{L \to \infty} (A_L / B_L) \leq 1 $, i.e.\ that $A_{L}$ is asymptotically bounded
above by $B_{L}$.

\section{Mirror-removal algorithm}

Our analysis is based on the idea of starting from a configuration with density $\pp = 1$,
in which each lattice site is occupied by a mirror, and removing mirrors to produce a new configuration with density $\pp
< 1$.  We will analyse how the quantities defined above change during this process, and we will 
obtain an upper bound for the total length of open paths.

\begin{figure}[h]
 \centering
 \includegraphics[scale=0.3]{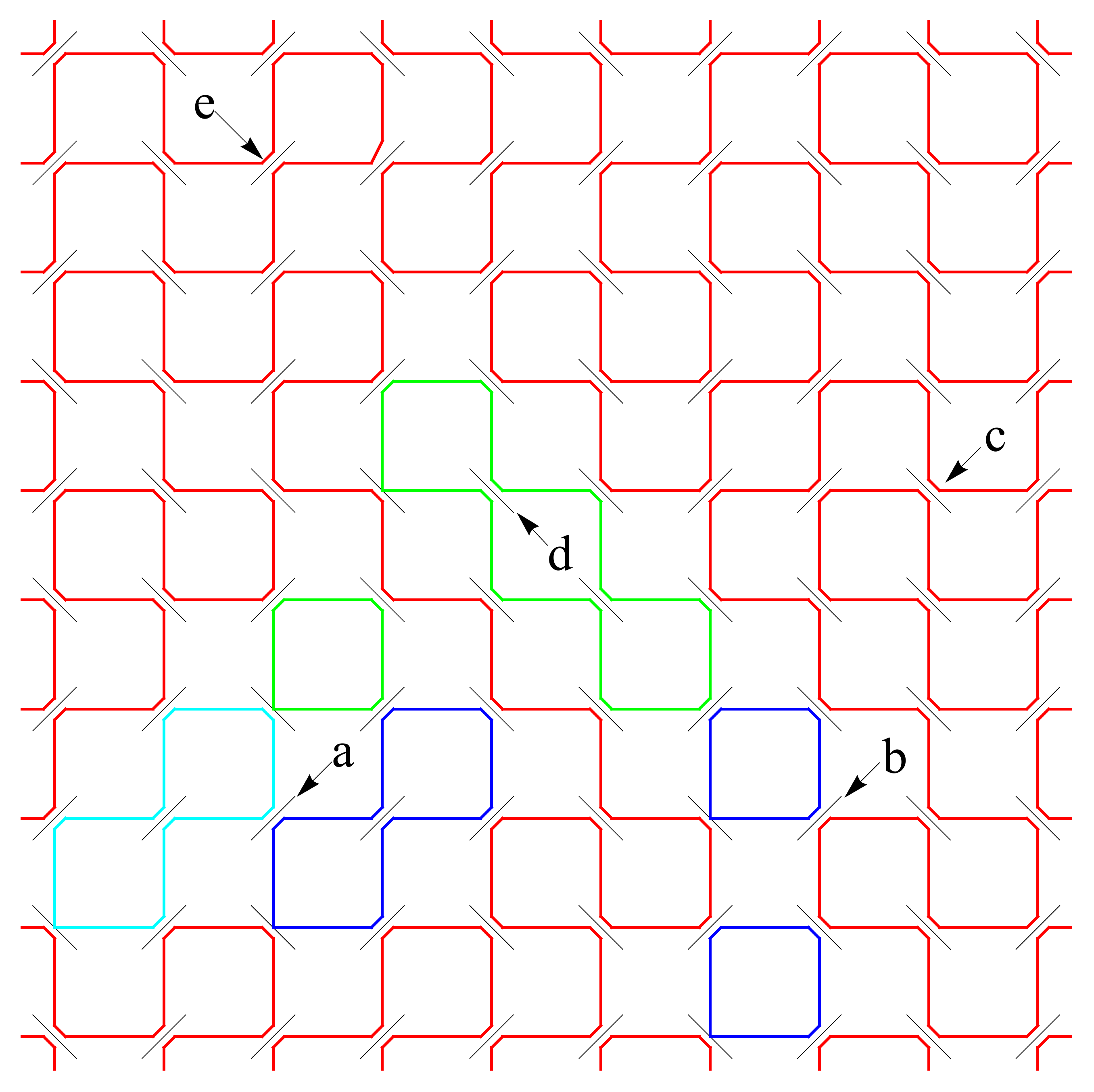}
 \caption{Trajectories in a $10 \times 10$ box. Open paths (those that leave the  box) are shown in red, 
 and touching closed paths are shown in different colours. Labels indicate the different possible
 mirror types: (a) separating two distinct closed paths; (b) separating an open and a closed path; (c) separating two distinct open paths; (d) embedded in a single closed path; and
(e) embedded in a single open path.}
 \label{fig:example_trajs}
\end{figure}

\begin{figure}[h]
\centering
\noindent\begin{tabular}{@{\hspace{0.0em}}c@{\hspace{1.5em}}c@{\hspace{0.0em}}}
\includegraphics[scale=0.35]{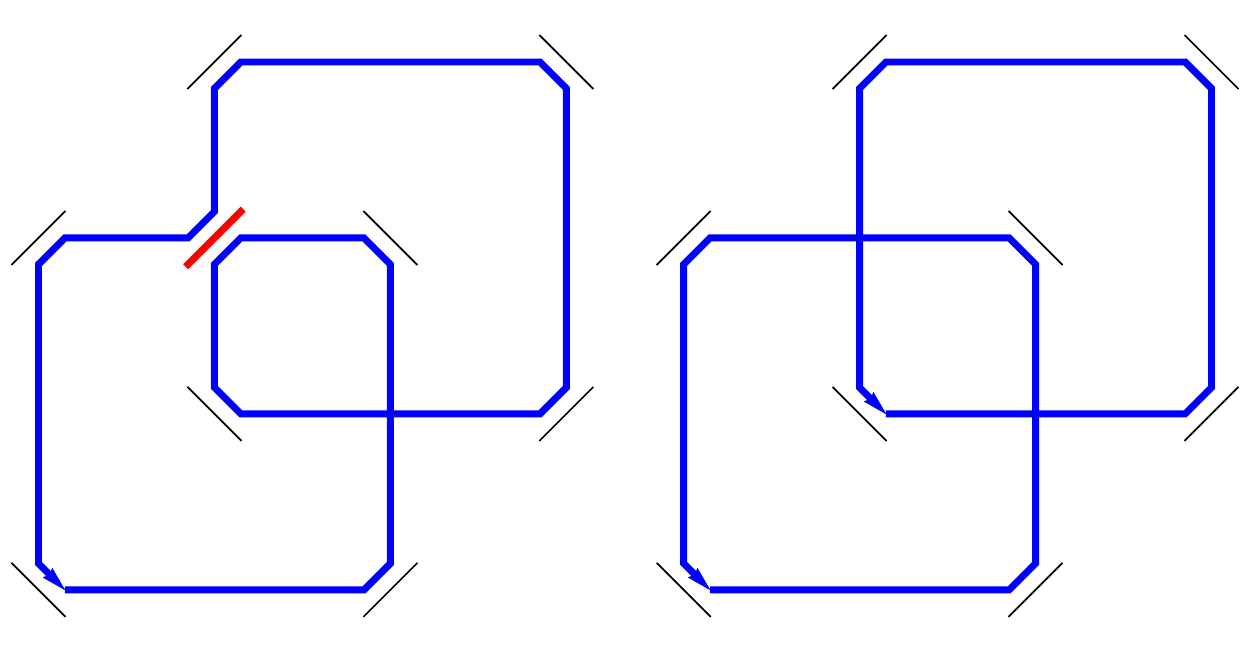} &
  \includegraphics[scale=0.35]{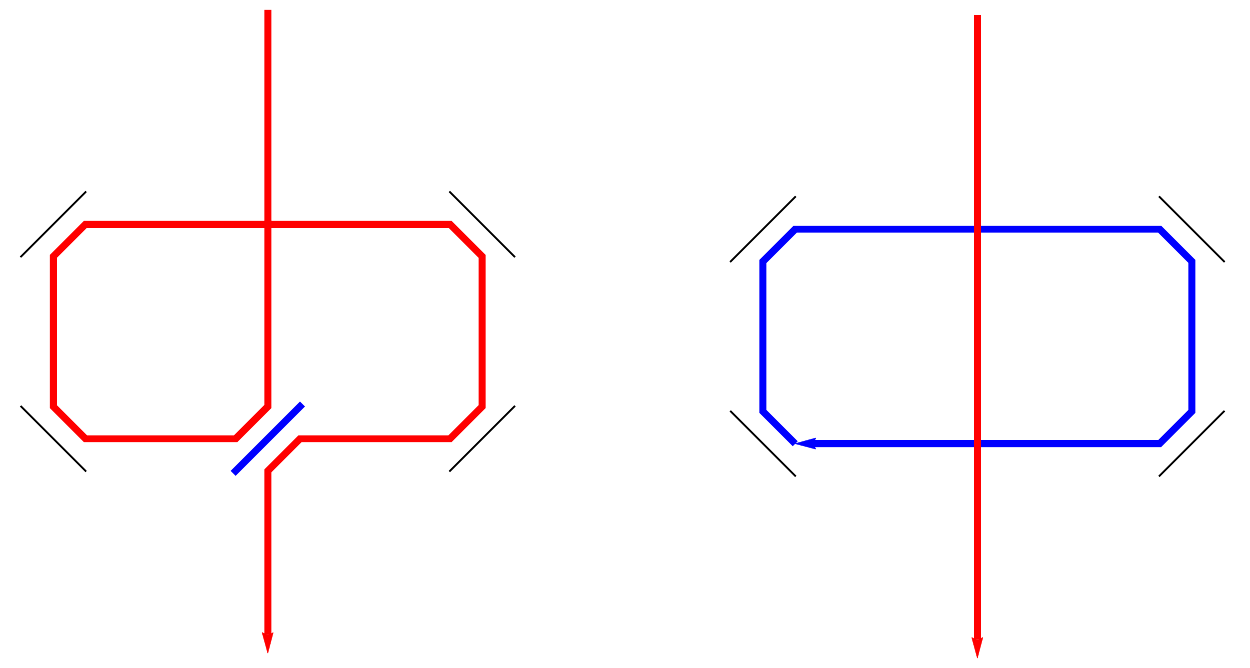}  \\
  (a) & (b)
\end{tabular}
 \caption{Creation of a closed path after removing a mirror. Red arrows represent
open paths; blue arrows represent close paths. (a) Two disjoint (but crossing) closed paths
generated by removing a mirror that separates a closed path from itself.
 (b) A closed path crossed by an open path, generated by removing a mirror that separates an open path
 from itself.}
 \label{fig:example_trajs2}
\end{figure}

Consider the removal of a single mirror. In either the finite or infinite version
of the model, one of the following possibilities occurs: (i) two distinct closed paths join together,
forming a single closed path with the same total length; (ii) two distinct open paths
exchange parts of their paths,
maintaining the same total open length; (iii) an open path joins with a closed
path, incorporating the closed path into the open path; (iv) the chosen mirror separates
two parts of the same
path (either open or closed). In case (iv), there are two possibilities: (a) the type and
length of the path are unaffected, but the order in which edges are visited changes; or
(b) the path is divided into a closed path and another path with the same type as the
original path before removing the mirror (closed or open).
The possible situations are illustrated in figure~\ref{fig:example_trajs} [cases (i)--(iv)(a)] and
figure~\ref{fig:example_trajs2} [case (iv)(b)].

Our algorithm proceeds by removing the required fraction of mirrors from the $\pp=1$ configuration (all vertices occupied by mirrors), in a certain order, in order to produce a configuration with $\pp<1$; any configuration with $\pp < 1$ may be produced in this way. The algorithm is as follows:
%

\begin{enumerate}
\item[(1)] Select, uniformly at random, the sites in the $L \times L$ box at which mirrors will be removed, each with probability $1-p$, independently of the rest; call this set $\mathcal{M}$.
\item[(2)] Remove, one by one, those mirrors in $\mathcal{M}$ that separate two \emph{distinct} closed paths.
\item[(3)] Remove those mirrors in $\mathcal{M}$ that join a closed path with itself. 
\item[(4)] Remove the mirrors in $\mathcal{M}$ that connect an open path with a closed path.
\item[(5)] Remove the remaining mirrors in $\mathcal{M}$, i.e.,
those that connect an open path with itself or with a different open path. 
\end{enumerate}

Note that since step (3) does not modify the total length of open paths, while step (5) can only reduce this length by producing closed paths (see figure~\ref{fig:example_trajs2}), we do not need to take them into account to obtain an upper bound.

\subsection{Joining distinct closed paths}
First consider step (2), in which mirrors that join two distinct closed paths are removed,
 creating a single closed path. 
Here, an order must be chosen in which to remove this subset of mirrors. When each mirror 
is checked, it is removed
only if it \emph{currently} separates two \emph{distinct} closed paths.
The exact configuration of closed paths obtained at the end of this step thus depends 
on the order in which these mirrors are removed.
Nonetheless, all the different configurations obtained, by removing these mirrors in any
order, have the same
number of closed paths, which we denote by $N_{L}(\pp)$, the same length $\CC$  (since the
total length of closed paths
has not changed in the process), and the same number of mirrors removed,  which we denote
by $\rr_{L}(\pp)$.

Each of the mirrors removed in this step reduces the number of closed paths by $1$. Thus
\begin{equation}
N_{L}(\pp) = N_{L}-\rr_{L}(\pp),
\label{eq:Mrho}
\end{equation}
where $N_{L} := N_{L}(1)$ is the number of closed paths when $p=1$. Note that
$\rr_{L}(\pp)$  is bounded above by $N_{L} - 1$.

We will argue that  $N_L(\pp) \gtrsim g(\pp) \, N_L$, where $g(\pp)$ is a positive function
of $\pp$ and $N_{L} := N_{L}(1)$ is the number of closed paths when $p=1$. 
Consider the closed paths with exactly four edges, which we call ``square paths'', and in particular those square paths that are not in contact with another square path. 
There is a positive probability that a given bond belongs to one of these paths; call this probability $\rho$. Since these square paths have four vertices, and none of them belongs to another square path, there are exactly 4 independent mirrors for each of these paths. Thus, the probability that one of the mirrors that belong to a given one of these closed paths is removed is $\pp^4$, and hence, the number of these closed paths after step (2) is bounded below by $\pp^4  \rho L^2/2$, so that $N_L(\pp) \gtrsim \pp^4  \, \rho \, L^2/2$. Since $N_L/L^2 \sim K \simeq 0.1$ (as measured numerically), for any $\pp > 0$ we have
%
%
\begin{equation}
N_{L}(\pp) \gtrsim \frac{ \pp^4 \rho K}{2}  N_L =: g(\pp) \, N_L,
\label{eq:N_L}
\end{equation}
with $g(\pp) := \frac{ \pp^4 \rho K}{2} >0$.

\subsection{Joining closed and open paths}

In step (4), we remove those mirrors that connect an open path with a closed path.
In this case, the closed path is amalgamated into the open path, so that the total open 
length grows by the length of this closed path.

\begin{figure}[t]
\begin{center}
\includegraphics*[scale=0.8]{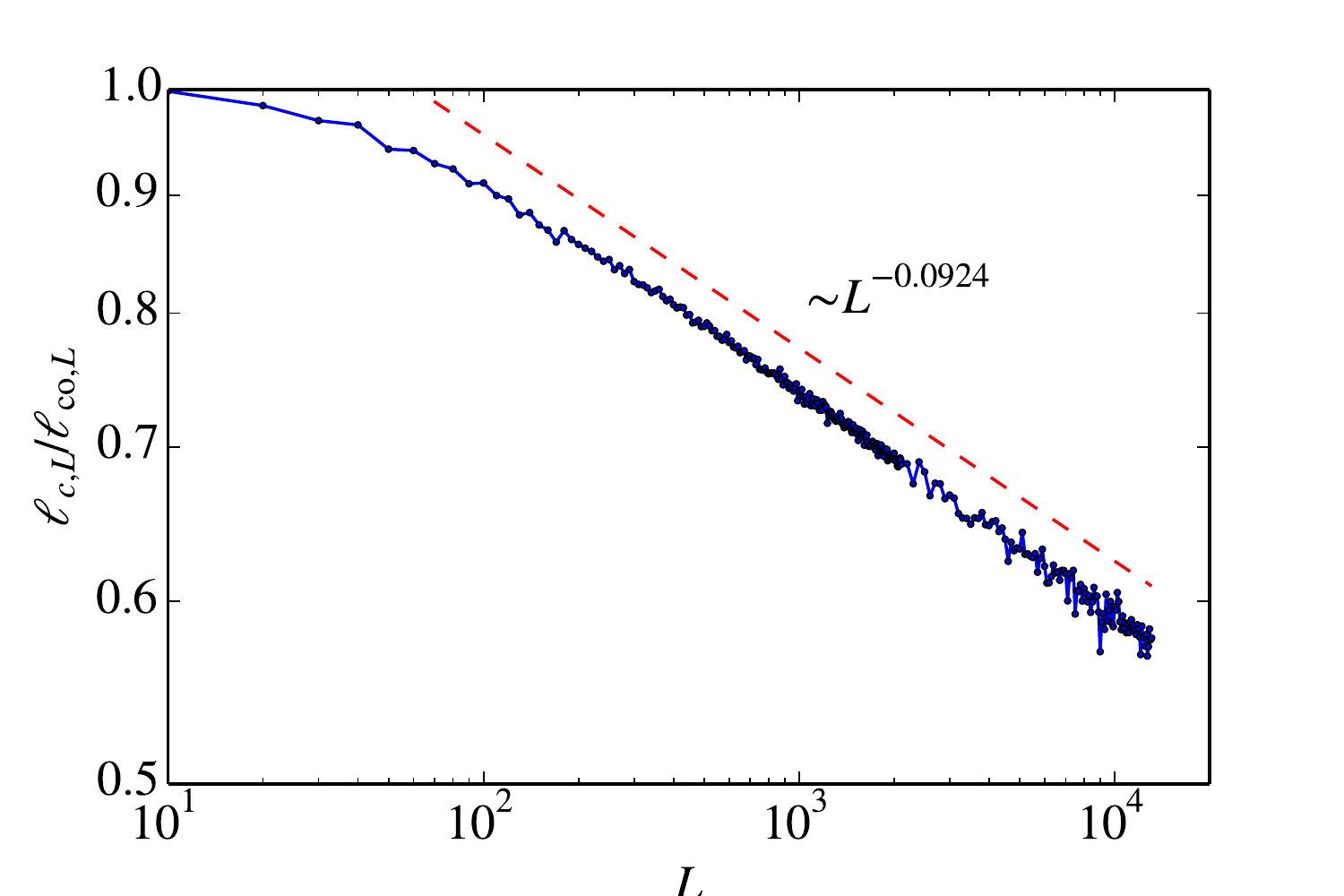}
\caption{Ratio $\lc / \lco$ of mean closed path length to mean length of those closed paths that touch open ones (double logarithmic scale). Each data point is averaged over at least 10 runs.}
\label{fig:types_of_closed_path}
\end{center}
\end{figure}

To estimate this increase, we use that the expected increase in the total open length is  
the \emph{mean} length of those closed paths that touch at least one open path; this mean
length we denote by $\lco$. 
To use this, we relate  it to the mean length of \emph{all} closed paths,
$\lc := \CC / N_{L}(\pp) \sim \CC / [N_{L} \, g(\pp)]$. (Note that the closed paths
referred  to are those \emph{after} the removal step (2).)
Figure~\ref{fig:types_of_closed_path} shows that for $\pp=1$, we have
\begin{equation}
\frac{\lc}{\lco} \sim C' \, L^{-\beta} \qquad \mathrm{with} \quad \beta \simeq 0.0924,
\label{eq:beta}
\end{equation}
and a constant $C'$.
(The efficient algorithm used for this calculation will be described in a separate
publication.)

\begin{figure}[t]
\begin{center}
\includegraphics*[scale=0.8]{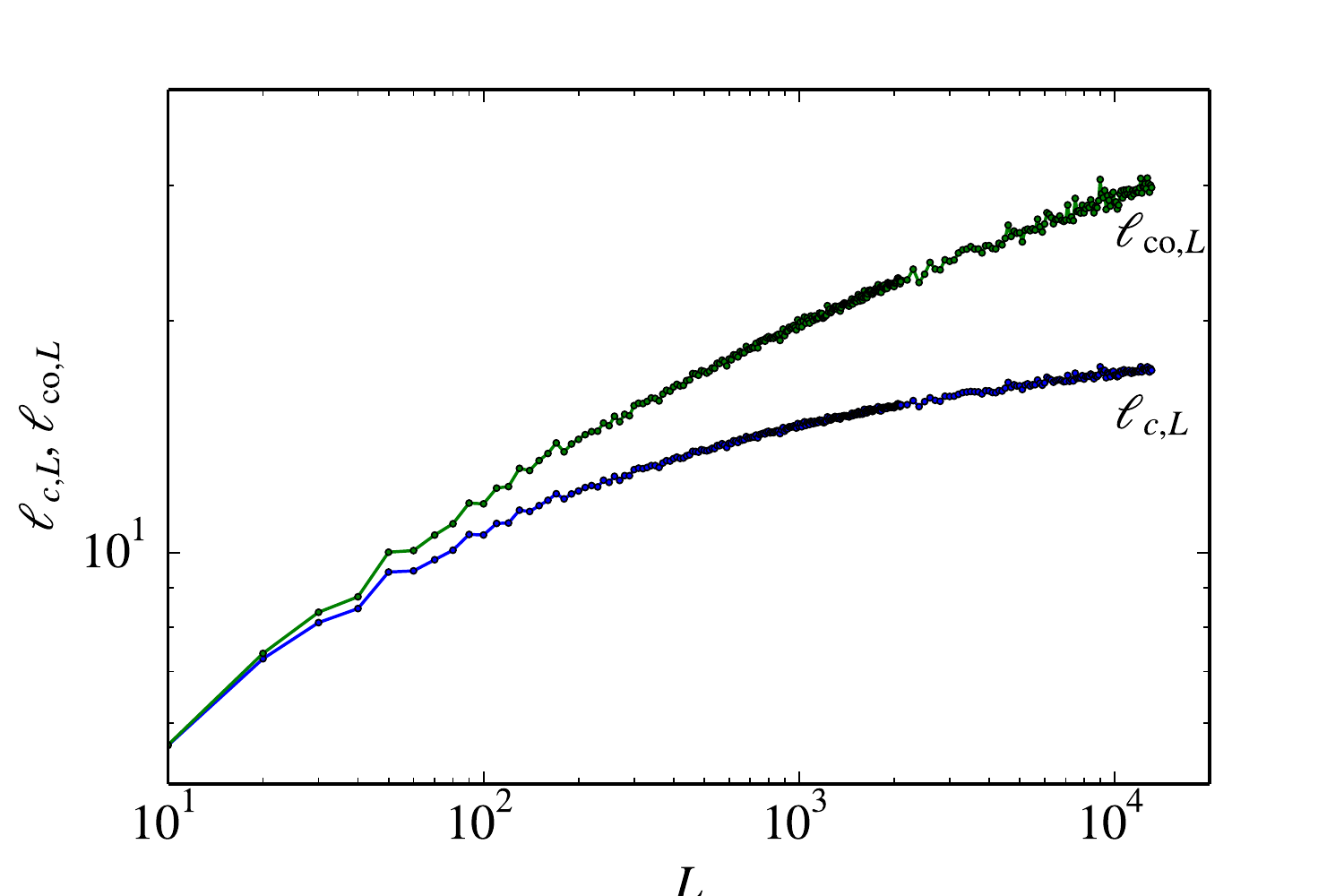}
\caption{Mean closed path length, $\lc$, and mean length of closed
paths that touch open ones, $\lco$  as a function of $L$ (double logarithmic scale).
}
\label{fig:ell-Cl}
\end{center}
\end{figure}

It is an open question whether this exponent $\beta$ is 
related with known quantities in percolation, as occurs with the exponent
$\alpha$. One possibility is the fractal dimension of the 2D percolation cluster, $d_f=91/48$ \cite{Sahimi}.
In figure~\ref{fig:ell-Cl} we plot separately $\lc$ and $\lco$ as functions of $L$. Although neither of these two
quantities appears to follow a power law, their ratio does. 
For sufficiently large values of $L$, we have that $\lc(L)$ may be estimated as $\lc \sim A \, (1-L^{-1/4})$,
with $A \simeq 20$. This result
is obtained by using that $\OO / 2L^{2}\sim L^{-1/4} $ (see figure
\ref{fig:open_length}), the definition $\lc := \CC / N_{L}$ for $\pp=1$, and the
fact that $N_{L} \sim K \, L^2$, where $K$ is a constant.

For other values of $\pp$, we will assume that
$\frac{\lc(\pp)}{\lco(\pp)} \sim L^{-\beta(\pp)}$. 
We wish to show that $\beta(\pp) \le \beta$ for all $0 < p \le 1$.
We have
\begin{equation}
\frac{\lc(\pp)}{\lco(\pp)} = \frac{\CC/N_L(\pp)}{\CCC(\pp)/\Nc(\pp)} =
\frac{[\CCC(\pp)+\CCR(\pp)]/N_L(\pp)}{\CCC(\pp) / \Nc(\pp)}
 \sim L^{-\beta(\pp)},
 \label{eq:talacha1}
\end{equation}
where $\CCC(\pp)$ is the length of the closed paths that touch an open path, $\CCR(\pp)$
is the length of the other closed paths and $\Nc(\pp)$ is the number of closed paths
touching open paths.
After reducing equation~\eqref{eq:talacha1}, we obtain
\begin{equation}
\frac{\lc(\pp)}{\lco(\pp)} =\left[ 1+ \frac{\CCR(\pp)}{\CCC(\pp)} \right] \,
\frac{\Nc(\pp)}{N_L(\pp)}
 \sim L^{-\beta(\pp)}.
\end{equation}

Suppose that the worst
case occurs: $\frac{\CCR(\pp)}{\CCC(\pp)} \sim 0$, so that 
$\frac{\Nc(\pp)}{N_L(\pp)} \sim L^{-\beta(\pp)}$. 
Using similar arguments to those used above to show  $N_L(\pp) \sim g(\pp) N_L$, we have 
$
\Nc(\pp) \gtrsim g_1(\pp) \Nc$, where $g_1(\pp)$ is a positive function of $\pp$. 
Thus, 
$\frac{\Nc(\pp)}{N_L(\pp)} \gtrsim f_1(\pp) \frac{\Nc}{N_L} \sim f_1(\pp) L^{-\beta}$, where $f_1(\pp)$ is again a positive function of $\pp$. Thus,
$0 \leq \beta(\pp) \leq \beta$.

%
%
%

The mean number of mirrors removed that join an open and a closed path 
is bounded above by $(1-\pp) \, 2L^{2} \, \Pv{o}{c}$. (Note that some mirrors join closed paths with themselves.)
Thus we have
\begin{equation}
\OOp{\pp} \lesssim \OO+(1-\pp) \, 2L^{2} \, \Pv{o}{c} \, \lco(\pp).
\label{eq:Op}
\end{equation}
(Here we have assumed that each joined closed path may be treated independently of the others, so that the total length increase is just given by the mean length increase times the number of such closed paths.)

Dividing equation~\eqref{eq:Op} by $2L^{2}$ and  using $\OO / 2L^{2} \lesssim C \, L^{-\alpha}$, we have
\begin{equation}
\frac{\OOp{\pp}}{2L^{2}} \lesssim C \, L^{-\alpha} + (1-\pp)  \Pv{o}{c} \, \lco(p).
\end{equation}
Now, using  $\lco(p) \sim  C' \, L^{\beta(p)} \lc \lesssim C' \, L^{\beta} \lc$ and $\lc =
\CC / N_{L}(\pp)$
and the bound 
$N_{L}(\pp) \gtrsim g(\pp) \, N_{L}$, we obtain
\begin{equation}
\frac{\OOp{\pp}}{2L^{2}} \lesssim C\, L^{-\alpha} + \frac{(1-\pp)}{g(\pp)} \, \Pv{o}{c} \, 
\frac{\CC}{ N_{L}} \, C' \, L^{\beta}.
\label{eq:master}
\end{equation}

Now, since $\Pv{o}{c} \lesssim  2 \Pop$, 
we have
\begin{equation}
\Pv{o}{c} \, \frac{\CC}{ N_{L} } 
\lesssim
2 \Pop \, \frac{\CC}{ N_{L}}
=
2 \frac{\OO} {N_{L}} \frac{\CC}{2L^{2}}
\lesssim 
2 \frac{\OO}{  N_{L} },
\end{equation}
since
$\Pop = \OO / 2L^{2}$ and
$\CC / 2L^{2} \rightarrow 1$.
Thus, equation~\eqref{eq:master} gives
\begin{equation}
\frac{\OOp{\pp}}{2L^{2}} \lesssim C\, L^{-\alpha} + \frac{2 (1-\pp)} {g(\pp)} \, C' \, 
\frac{\OO}{N_{L}} \, L^{\beta}.
\end{equation}
Recalling that $N_{L} \gtrsim K L^{2}$, we finally have
\begin{equation}
\frac{\OOp{\pp}}{2L^{2}} \lesssim C\, L^{-\alpha} + \frac{2 (1-\pp) C'} {K g(\pp)}  \, 
\frac{\OO}{L^{2}} \, L^{\beta}, 
\end{equation}
and hence
\begin{equation}
\frac{\OOp{\pp}}{2L^{2}} \lesssim C\, L^{-\alpha} + \frac{2 (1-\pp) C'} {K g(\pp)} \, L^{\beta - \alpha}.
\end{equation}

Since our numerical results show that $\beta < \alpha$, we conclude that
$\frac{\OOp{\pp}}{2L^{2}} \sim 0$, that is, the density of open paths tends to $0$, or equivalently
the density of closed paths tends to $1$, as the box size goes to infinity,
for \emph{any} mirror density $p$.

\section{Conclusions}

In this paper, we have shown, using input from numerical calculations, that in the limit
of infinite system size the density of open paths in the Lorentz mirror model tends to
$0$ for any mirror probability $0 < p \le 1$. This is a weaker form of the conjecture that with probability $1$ all trajectories
close.

Our argument reduces the problem to the analysis of the two exponents $\alpha$ and
$\beta$,
which we calculated numerically via efficient simulations. Clearly, it is important to calculate $\beta$
analytically.  Furthermore, our analysis has dealt only with mean quantities; it is also necessary
to study the probability distributions of the corresponding random variables.

The authors thank L.~Bunimovich, E.~Crane, T.~LaGatta and R. Díaz for useful discussions
and comments on earlier drafts, and an anonymous referee for useful suggestions which improved the manuscript. Financial support from CONACYT grant CB-101246 and
DGAPA-UNAM PAPIIT grants IN116212 and IN117214 is acknowledged.


\begin{thebibliography}{10}
\providecommand{\url}[1]{{#1}}
\providecommand{\urlprefix}{URL }
\expandafter\ifx\csname urlstyle\endcsname\relax
  \providecommand{\doi}[1]{DOI~\discretionary{}{}{}#1}\else
  \providecommand{\doi}{DOI~\discretionary{}{}{}\begingroup
  \urlstyle{rm}\Url}\fi

\bibitem{Acedo}
Acedo, L., Santos, A.: Diffusion in lattice {L}orentz gases with mixtures of
  point scatteres.
\newblock Phys. Rev. E \textbf{50}, 4577 (1994)

\bibitem{Beijeren}
van Beijeren, H., Ernst, M.H.: Diffusion in {L}orentz lattice gas automata with
  backscattering.
\newblock J. Stat. Phys. \textbf{70}, 793 (1993)

\bibitem{164}
Bianca, C.: On the existence of periodic orbits in nonequilibrium {E}hrenfest
  gas.
\newblock In: International Mathematical Forum, vol.~7, pp. 221--232 (2012)

\bibitem{1}
Bruin, C.: A computer experiment on diffusion in the {L}orentz gas.
\newblock Physica \textbf{72(2)}, 261--286 (1974)

\bibitem{Bunimovich}
Bunimovich, L.A., Troubetzkoy, S.E.: Recurrence properties of {L}orentz lattice
  gas cellular automata.
\newblock J. Stat. Phys. \textbf{67}(1), 289--302 (1992)

\bibitem{Bunimovich2}
Bunimovich, L.A., Troubetzkoy, S.E.: Topological dynamics of flipping {L}orentz
  lattice gas models.
\newblock J. Stat. Phys. \textbf{72}, 297 (1993)

\bibitem{Cohen2}
Cohen, E.: New types of diffusion in lattice gas cellular automata.
\newblock In: M.~Mareschal, B.~Holian (eds.) Microscopic Simulations of Complex
  Hydrodynamic Phenomena, \emph{NATO ASI Series}, vol. 292, pp. 137--152.
  Springer US (1992)

\bibitem{Wang}
Cohen, E., Wang, F.: New results for diffusion in {L}orentz lattice gas
  cellular automata.
\newblock J. Stat. Phys. \textbf{81}, 445 (1995)

\bibitem{Cohen1}
Cohen, E., Wang, F.: Diffusion and propagation in {L}orentz lattice gases.
\newblock Fields Institute Communications \textbf{6}, 43 (1996)

\bibitem{M-H-Ernst2}
Ernst, M.H., van Velzen, G.A.: Lattice {L}orentz gas.
\newblock J. Phys. A: Math. Gen. \textbf{22}, 4611 (1989)

\bibitem{M-H-Ernst1}
Ernst, M.H., van Velzen, G.A.: Long-time tails in lattice {L}orentz gases.
\newblock J. Stat. Phys. \textbf{57}, 455 (1989)

\bibitem{10}
Grimmett, G.: Percolation and disordered systems.
\newblock In: Lectures on Probability Theory and Statistics, pp. 153--300.
  Springer (1997)

\bibitem{181}
Grimmett, G.: Percolation, 2nd edn.
\newblock Springer-Verlag, New York (1999)

\bibitem{XP-Kong3}
Kong, X.P., Cohen, E.G.D.: Diffusion and propagation in triangular {L}orentz
  lattice gas cellular automata.
\newblock J. Stat. Phys. \textbf{62}, 737 (1991)

\bibitem{XP-Kong1}
Kong, X.P., Cohen, E.G.D.: {L}orentz lattice gases, abnormal diffusion, and
  polymer statistics.
\newblock J. Stat. Phys. \textbf{62}, 1153 (1991)

\bibitem{GADY-KOZMA}
Kozma, G., Sidoravicius, V.: Lower bound for the escape probability in the
  {L}orentz mirror model on the lattice.
\newblock arXiv:1311.7437 [math.PR] pp. 1--2 (2013)

\bibitem{ata-david}
Kraemer, A.S., Sanders, D.P.: Embedding quasicrystals in a periodic cell:
  Dynamics in quasiperiodic structures.
\newblock Phys. Rev. Lett. \textbf{111}(12), 5501 (2013)

\bibitem{5}
Machta, J., Moore, S.M.: Diffusion and long-time tails in the overlapping
  {L}orentz gas.
\newblock Phys. Rev. A \textbf{32}, 3164--3167 (1985)

\bibitem{6}
Machta, J., Zwanzig, R.: Diffusion in a periodic {L}orentz gas.
\newblock Phys. Rev. Lett. \textbf{50}, 1959–1962 (1983)

\bibitem{7}
Moran, B., Hoover, W.G.: Diffusion in a periodic {L}orentz gas.
\newblock J. Stat. Phys. \textbf{48}, 709--726 (1987)

\bibitem{Quas}
Quas, A.N.: Infinite paths in a {L}orentz lattice gas model.
\newblock Prob. Theor. Rel. Fields \textbf{114}(2), 229--244 (1999)

\bibitem{Ruijgrok}
Ruijgrok, T.W., Cohen, E.: Deterministic lattice gas models.
\newblock Phys. Lett. A \textbf{133}(7), 415--418 (1988)

\bibitem{Sahimi}
Sahimi, M.: Applications of percolation theory.
\newblock Taylor and Francis (2009)

\bibitem{Wang2}
Wang, F., Cohen, E.: Diffusion in {L}orentz lattice gas cellular automata: The
  honeycomb and quasi-lattices compared with the square and triangular
  lattices.
\newblock J. Stat. Phys. \textbf{81}, 467 (1995)

\bibitem{XP-Kong2}
X.P.~Kong, E.C.: A kinetic theorist's look at lattice gas cellular automata.
\newblock Physica D \textbf{47}, 9--18 (1991)

\bibitem{Ziff2}
Ziff, R.M.: Hull-generation walks.
\newblock Phys. D \textbf{38}, 377--383 (1989)

\bibitem{ziff}
Ziff, R.M., Kong, X., Cohen, E.: {L}orentz lattice-gas and kinetic-walk model.
\newblock Phys. Rev. A \textbf{44}(4), 2410 (1991)

\end{thebibliography}
\end{document}